# Hydrodynamic Phonon Transport: Past, Present, and Prospect

[1,2]Sangyeop Lee*, [1]Xun Li

[1]Department of Mechanical Engineering and Materials Science, University of Pittsburgh, Pennsylvania, 15261, USA

[2]Department of Physics and Astronomy, University of Pittsburgh, Pennsylvania, 15261, USA

*sylee@pitt.edu

**Abstract**

The hydrodynamic phonon transport was studied several decades ago for verifying the quantum theory of lattice thermal transport. Recent prediction of significant hydrodynamic phonon transport in graphitic materials shows its practical importance for high thermal conductivity materials and brought a renewed attention. As the study on this topic has been inactive to some extent for several decades, we aim at providing a brief overview of the past studies as well as very recent studies. The topics we discuss in this chapter include the collective motion of phonons, several approaches to solve the Peierls-Boltzmann transport equation for hydrodynamic phonon transport, the role of normal scattering for thermal resistance, and the propagation of second sound. Then, we close this chapter with our perspectives for the future studies and the practical implication of hydrodynamic phonon transport.

# I. Introduction

The transport of phonons, a major heat carrier in non-metallic solids, has been usually described by the diffusive limit since the Fourier's law was suggested 200 years ago. The Fourier's law has a simple form that correlates thermodynamic driving force (i.e. temperature gradient, $-\nabla T$) and the resulting heat flux ($q''$):

$$\frac{1}{\kappa}q'' = -\nabla T \qquad (1)$$

This empirical law shows that there is always a damping coefficient, $1/\kappa$, involved in the transport phenomena. The $1/\kappa$ is thermal resistance which determines the extent of damping in heat flow and the resulting heat flux at a given temperature gradient. However, such a damping effect is not observed in fluid flow although both phonons and molecules are well-described by the same Boltzmann transport theory. Also, they have similar thermodynamic driving forces; molecules are driven by pressure gradient like phonons are driven by temperature gradient. Assuming an infinitely large domain to exclude any effect from boundary, the molecular flow at macroscale can be described by the Euler's equation:

$$\frac{D(\rho \mathbf{u})}{Dt} = -\nabla p \qquad (2)$$

where $\rho$ and $\mathbf{u}$ are the density and velocity of fluid element. With the Lagrangian coordinate, Eq. (2) shows the acceleration of molecules under the pressure gradient ($-\nabla p$) without any damping effect. This is the thermodynamic limit where the entropy generation is zero.

Now one may ask a question – why does phonon flow described by the Fourier's law exhibit a damping effect while molecular flow does not? Interestingly, Nernst speculated a century ago that heat in high thermal conductivity materials may have inertia like fluid [1]. The different behaviors of damping in molecular and phonon flows can be associated with the difference in scattering processes of those two particles in terms of the momentum conservation. For molecular flow, total momentum of molecules is always conserved upon molecule-molecule scattering. Therefore, inter-molecular scattering itself cannot cease the given molecular flow. For phonon flow, however, the total momentum of phonons is not always conserved upon phonon-phonon scattering. There are two different scattering mechanisms regarding the momentum conservation: normal and umklapp scattering (hereafter N-scattering and U-scattering, respectively), suggested by Peierls [2]. As shown in Fig. 1(a), the N-scattering involves phonon states with small wavevectors and the total momentum of phonon particles is conserved ($\mathbf{q}_1 + \mathbf{q}_2 = \mathbf{q}_3$) like inter-molecular scattering case. However, for U-scattering, the total momentum of phonon particles is not conserved. As a result, the phonon propagation direction is reversed upon U-scattering, thus directly causing thermal resistance. Phonon scattering by impurities shown in Fig. 1c also directly causes thermal resistance as it does not conserve total momentum. Hereafter R-scattering refers to combined U- and impurity-scattering. In most materials at room temperature, N-scattering is weak compared to R-scattering, leading to the large damping effect of heat flow in solid materials.

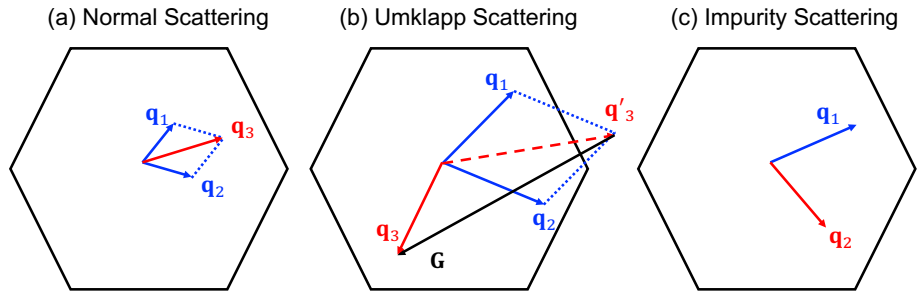

Fig. 1. Schematic of N-scattering, U-scattering, and impurity scattering in the reciprocal space. The hexagon represents the first Brillouin zone.

We have assumed the infinitely large domain to compare the intrinsic damping of phonon flow and molecular flow. However, all solid materials have finite size, introducing phonon-boundary scattering. In most cases, the phonon-boundary scattering is diffuse boundary scattering rather than specular boundary scattering. The three types of phonon scattering (i.e., diffuse boundary scattering, N-scattering, and R-scattering) influence phonon transport in different ways and thus there exist three regimes of phonon transport – ballistic, hydrodynamic, and diffusive regimes schematically shown in Fig. 2 – depending on the dominant type of scattering mechanisms. Those three regimes occur in different ranges of temperature. The ballistic regime occurs at low temperature where internal phonon scattering is much weaker than phonon-boundary scattering. Therefore, the phonon transport is limited by the diffuse boundary scattering and the thermal resistance is determined by the size and shape of samples. As temperature increases, the internal phonon scattering starts to play a role in the transport process. At sufficiently low temperature, majority of internal phonon scattering is N-scattering as phonon states with large wavevectors cannot be occupied. Because of the momentum conserving nature of N-scattering, the resulting phonon transport is similar to fluid flow and thus called hydrodynamic phonon transport. Fig. 2(b) shows the schematic of heat flux profile, similar to the molecular Poiseuille flow. When temperature increases further, U-scattering becomes significant and the thermal resistance is due to the direct momentum destruction by U-scattering. As U-scattering occurs in any location, the heat flux has a spatially uniform profile as shown in Fig. 2(c).

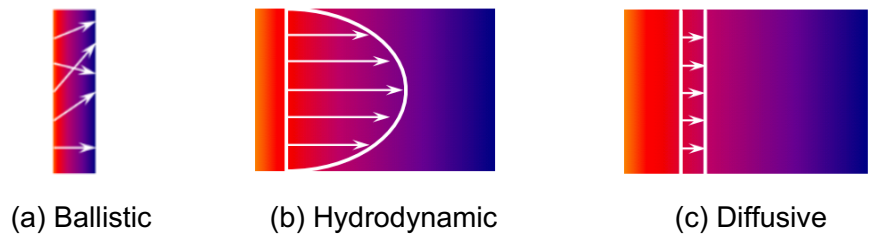

(a) Ballistic        (b) Hydrodynamic        (c) Diffusive

Fig. 2. Schematic of phonon flux profile in ballistic, hydrodynamic, and diffusive regimes [3]

The N- and U-scattering for phonons, since suggested by Peierls around a century ago [2], have been a foundation for the quantum theory of thermal transport in solids. Although the concept of N- and U-scattering were well accepted, the direct confirmation of N-scattering was still lacking. This led to the theoretical [4-10] and experimental efforts [11-15] for the prediction and observation of hydrodynamic phonon transport, namely phonon Poiseuille flow and second sound which will be discussed later in more detail. The phonon Poiseuille flow was first measured in solid He at the temperature range of 0.6 to 1.0 K [11]. The second sound was measured in solid $^3$He at 0.5 K [12], in NaF at around 15 K [13, 14], and in Bi at 2 K [15]. Those experimental observations combined with the theoretical studies directly confirm the N-scattering for phonons and show remarkably different effects of N- and U-scattering on thermal transport. This was regarded as "one of the great triumphs of the theory of lattice vibrations" [16].

Despite of the confirmation of hydrodynamic phonon transport, the study on this topic has been inactive for the recent several decades. As can be seen from the previous measurements, the hydrodynamic phonon transport was observed at very low and narrow temperature ranges and thus considered not relevant to practical applications. The conditions for hydrodynamic phonon transport are stringent because it is rare to satisfy the weak U-scattering and strong N-scattering at the same time. The U-scattering can be easily suppressed if temperature is much lowered than the Debye temperature so as to limit the phonon population to small wavevector states. However, if the temperature is lowered, there is not enough N-scattering events and the transport easily becomes ballistic. Thus, for hydrodynamic phonon transport to be significant, a material should exhibit a high Debye temperature and large anharmonicity at the same time. This is not common; a material with a high Debye temperature like diamond usually exhibits small anharmonicity. The quality of sample is another issue as the impurity scattering is the momentum-destroying scattering and weakens the hydrodynamic features. It is interesting to note that NaF was chosen for the second sound experiments [13, 14, 17] because Na and F are naturally monoisotopic elements and thus at least isotope impurity does not exist.

The hydrodynamic phonon transport has recently received a renewed attention after first-principles-based studies predicted the significant hydrodynamic phonon transport in graphitic materials including single-wall carbon nanotubes (SWCNTs) [18], graphene [3, 19], and graphite [20]. Interestingly, those graphitic materials exhibit a high Debye temperature and large anharmonicity at the same time, leading to the strong N-scattering shown in Fig. 3 and the significant hydrodynamic phonon transport [3]. The light atomic mass of carbon and strong sp$^2$ bonding result in the high Debye temperature and weak U-scattering. Also, the flexural phonon modes from its layered atomistic structure are largely anharmonic for small wavevector states [21], leading to the strong N-scattering.

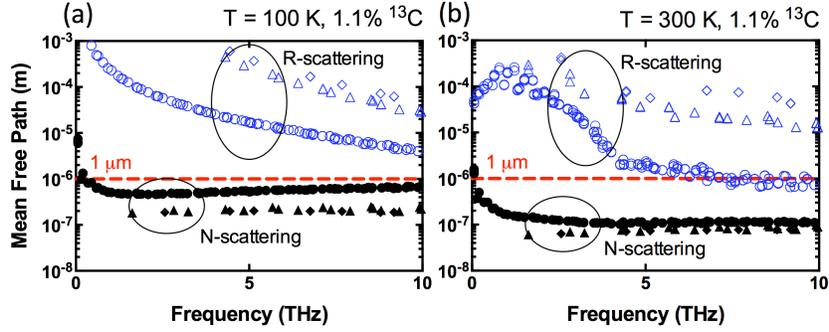

Fig. 3. The mean free paths of N- and R-scattering in suspended graphene at 100 and 300 K from first-principles calculation.

The primary objective of this book chapter is to provide a brief overview of basic concepts and recent studies of hydrodynamic phonon transport for those who have previously worked on ballistic and diffusive phonon transport. Other comprehensive review articles are available for advanced theoretical aspects [22, 23] and macroscopic governing equations of heat wave which is related to second sound [24]. This book chapter is organized as follows. Section II discusses the displaced Bose-Einstein distribution as an equilibrium distribution under N-scattering and collective hydrodynamic phonon flow. Section III summarizes the methods to solve the Peierls-Boltzmann transport equation for hydrodynamic and quasi-hydrodynamic phonon transport. Section IV provides our current understanding on the role of N-scattering for thermal resistance for various cases. Section V will review the theoretical and experimental studies of second sound. We then briefly discuss the future perspectives of phonon hydrodynamics in Section VI.

We also would like to mention that the term 'hydrodynamic phonon transport' has been used in a different context in recent publications [25-31]. Those studies used phonon hydrodynamic equations that were derived assuming strong N-scattering compared to U-scattering and thus have a term similar to the viscous term of the Navier-Stokes equation [7, 8, 32]. However, to avoid any confusion, the phenomena studied in those studies are quasi-ballistic phonon transport and do not require the strong N-scattering; the hydrodynamic equations were used to phenomenologically describe the quasi-ballistic transport. In this book chapter, we focus on the hydrodynamic phenomena of phonon transport due to strong N-scattering and do not discuss the phenomenological hydrodynamic description of quasi-ballistic phonon transport. For readers who are interested in the latter topic, a recent review article can provide a comprehensive summary [33].

## II. Collective Phonon Flow

One unique feature of hydrodynamic transport that can be distinguished from ballistic and diffusive regimes is the collective motion of particles. The term 'collective' is often used to describe different phenomena in solid-state physics. Here, we call the transport of particles is collective when the flux of particles can be represented by a single value of velocity regardless of their quantum states. As an example, let us assume that we are able to track the movements of all molecules in a small fluid element. Assuming strong molecule-molecule scattering and small pressure gradient for the well-defined local equilibrium condition, the molecules then follow the displaced Boltzmann distribution:

$$f_B^{disp} = \left(\frac{m}{2\pi k_B T}\right)^{3/2} \exp\left(-\frac{m|\mathbf{v}-\mathbf{u}|^2}{2k_B T}\right) \tag{3}$$

where $m$, $k_B$, and $T$ represent mass of a molecule, the Boltzmann constant and temperature, respectively. The $\mathbf{v}$ is the actual velocity of a molecule and $\mathbf{u}$ is the drift velocity. Note that the drift velocity is the same for all molecules regardless of their quantum states. Usually the actual velocity is much larger than the drift velocity, making the movement of each molecule looks like random. However, the small drift velocity causes a net flow of molecules. As a result, the fluid element containing many molecules that seemingly move along random direction can move with the drift velocity as a whole. Thus, we call the molecular transport collective in this case.

Likewise, phonon particles show the collective motion when the transport is hydrodynamic. The equilibrium distribution of phonons with N-scattering is the displaced Bose-Einstein distribution:

$$f^{disp} = \left[\exp\left(\frac{\hbar(\omega - \mathbf{q}\cdot\mathbf{u})}{k_B T}\right) - 1\right]^{-1} \tag{4}$$

where $\mathbf{q}$ and $\mathbf{u}$ are the phonon wavevector and drift velocity (or displacement), respectively. In most cases where the transport is non-hydrodynamic, $\mathbf{u}$ differs for each phonon mode. However, in hydrodynamic regime $\mathbf{u}$ is a constant for all phonon modes. The displaced distribution function can be linearized assuming a small displacement, i.e., $\mathbf{q}\cdot\mathbf{u} \ll \omega$,

$$f^{disp} \approx f^0 + \frac{\hbar}{k_B T} f^0 \left(f^0 + 1\right) \mathbf{q}\cdot\mathbf{u} \tag{5}$$

The fact that the displaced Bose-Einstein distribution function is the equilibrium distribution upon N-scattering can be shown with the Boltzmann's *H*-theorem [34]. For example, the rate of entropy generation upon coalescence three-phonon scattering is

$$\dot{S}_{scatt} \sim \sum_{ijk} \left(\phi_i + \phi_j - \phi_k\right)^2 P_{i,j}^k \tag{6}$$

where $P_{i,j}^k$ is the equilibrium transition rate of the coalescence process where the phonon particles at the state *i* and *j* are merged to the state *k*. A similar expression can be written for the decay process. The $\phi_i$ represents the deviation of distribution function from the stationary Bose-Einstein distribution $f_i^0$ (i.e., displaced Bose-Einstein distribution with zero displacement), and is defined as $\phi_i = (f_i - f_i^0)/(f_i^0(f_i^0+1))$. If the three phonon states exhibit the displaced Bose-Einstein distribution,

$$\phi_i + \phi_j - \phi_k = (\mathbf{q}_i + \mathbf{q}_j - \mathbf{q}_k) \cdot \mathbf{u} \tag{7}$$

Considering the momentum conservation of N-scattering, $\mathbf{q}_i + \mathbf{q}_j = \mathbf{q}_k$, the entropy generation in this case is zero, verifying that the displaced Bose-Einstein distribution is an equilibrium distribution under N-scattering. From Eq. (7), even U-scattering ($\mathbf{q}_i + \mathbf{q}_j = \mathbf{q}_k \pm \mathbf{G}_m$) does not generate any entropy if the reciprocal lattice vector $\mathbf{G}_m$ is orthogonal to $\mathbf{u}$. This was also shown through the simulation of second sound in a recent study [35].

Whether a certain scattering process is N- or U-scattering depends on the choice of the Brillouin zone, which may lead to confusion or misunderstanding about the role of N- and U-scattering on phonon transport. We would like to emphasize that the concept of momentum conservation for understanding the phonon transport is valid only when the crystal momentum is defined with the first Brillouin zone which is the Wigner-Seitz unit cell in reciprocal space. Otherwise, the displaced distribution function in Eq. (4) is incorrect and distinguishing N- and U-scattering based on the non-Wigner-Seitz unit cell is not meaningful.

With the linearized form of the displaced distribution function in Eq. (5), it is straight forward to show that the phonon particle flux $n_x''$ can be described by the single value of $\mathbf{u}$:

$$n_x'' = \frac{1}{NV}\sum_i v_x f^{\mathrm{disp}} = \left(\frac{1}{NV}\sum_i v_x \frac{\hbar}{k_B T} f^0(f^0+1)\mathbf{q}\right) \cdot \mathbf{u} \tag{8}$$

where $N$ and $V$ are the number of atoms and the volume of unit cell. Similarly, heat flux $q_x''$ is

$$q_x'' = \frac{1}{NV}\sum_i \hbar\omega v_x f^{\mathrm{disp}} = \left(\frac{1}{NV}\sum_i \hbar\omega v_x \frac{\hbar}{k_B T} f^0(f^0+1)\mathbf{q}\right) \cdot \mathbf{u} \tag{9}$$

It is noteworthy that both particle flux and heat flux are linearly proportional to the local drift velocity $\mathbf{u}$, representing the collective motion of phonon particles. The coefficients in the parenthesis are constants determined by phonon dispersion and temperature. The fact that single value $\mathbf{u}$ can describe the transport of all phonon particles is the basis for the macroscopic transport equation about $\mathbf{u}$ which will be discussed in the section III.

As U-scattering cannot be completely avoided, the actual phonon distribution deviates from the displaced Bose-Einstein distribution to some extent:

$$f_i \approx f_i^0 + \frac{\hbar}{k_B T} f_i^0(f_i^0+1)(\mathbf{q}_i \cdot \mathbf{u} + \delta_i) \tag{10}$$

where $\delta$ represents the deviation from the displaced Bose-Einstein distribution. It would be interesting to see how close the actual phonon distribution is to the displaced Bose-Einstein distribution in real materials in which hydrodynamic phonon transport is expected to be significant. In Fig. 4, we show the distribution function of phonon particles along the armchair direction in graphene at 100 K from the Peierls-Boltzmann transport equation (PBE) in an infinitely large sample case which will be discussed in the section III. In most cases where the transport is not hydrodynamic, $\delta_i$ in Eq. (10) is larger compared to the collective part $\mathbf{q}_i \cdot \mathbf{u}$ and thus $(f_i - f_i^0)/(f_i^0(f_i^0+1))$ is not linear to $q_{i,x}$. However, in graphene, $(f_i - f_i^0)/(f_i^0(f_i^0+1))$ is nearly linear to $q_{i,x}$ with a constant slope, representing the collective motion of phonon particles with the same displacement regardless of the phonon mode. Fig. 5 shows the contribution of the collective motion of phonon particles to total heat flux in (20,20) SWCNTs. At low temperature below 100 K, most of heat is carried by the collective motion of phonon particles and the contribution of collective motion gradually decreases with temperature due to U-scattering.

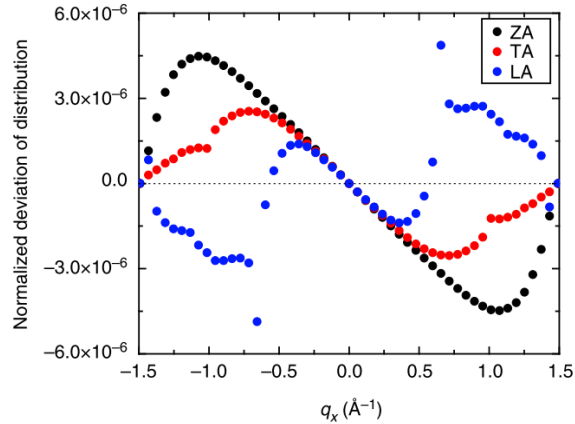

Fig. 4. Normalized deviational distribution, $(f_i - f_i^0)/(f_i^0(f_i^0+1))$, in an infinitely large graphene at 100 K [3].

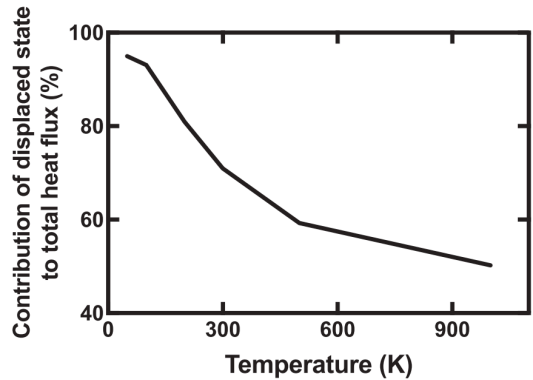

Fig. 5. Contribution of collective motion of phonon particles to total heat flux in (20,20) SWCNT with naturally occurring $^{13}$C isotope content (1.1%) [18].

## III. Peierls-Boltzmann Transport Equation

The phonon distribution is described by the PBE:

$$\frac{\partial f_i(t,\mathbf{x})}{\partial t} + \mathbf{v}_i \cdot \nabla f_i(t,\mathbf{x}) = \sum_j G_{ij} f_j^d \tag{11}$$

where $f_j^d$ is the deviational distribution function defined as $f_j - f_j^0$ and the $\mathbf{G}$ is the scattering matrix. The original form of the PBE is known to be difficult to solve. The advection and scattering terms are in differential and integral forms and the unknown, $f_i(t,\mathbf{x})$, is a function in many dimensions including time, real space, and reciprocal space domains. The equation has been often simplified assuming steady state, a constant temperature gradient in an infinitely large sample, and very small deviation from the equilibrium distribution:

$$\mathbf{v}_i \cdot \nabla T \frac{df_i^0}{dT} = \sum_j G_{ij} f_j^d \tag{12}$$

The differential advection term in Eq. (11) was replaced with the spatially homogenous term, $\mathbf{v}_i \cdot \nabla T \left( df_i^0/dT \right)$, by assuming the constant temperature gradient in an infinitely large sample. With these assumptions, the phonon distribution function is spatially homogenous except for the change due to temperature gradient. Then, the PBE could be simplified from the integro-differential equation to the homogenous integral equation which is relatively easier to solve. Recently developed *ab initio* framework of lattice dynamics made it possible to calculate the scattering matrix, $\mathbf{G}$, from first principles [37, 38]. Also, several numerical techniques such as the full iterative method [39, 40] and the variational method [41] were developed to solve the Eq. (12). Solving Eq. (12) with *ab initio* phonon dispersion and scattering matrix showed an excellent predictive power for thermal conductivity of bulk samples [42].

For the hydrodynamic regime, however, the assumption of spatially homogenous distribution function is not valid. As schematically shown in Fig. 2(b), the heat flux and phonon distribution largely depends on the location in real space, and the advection term, $\mathbf{v}_i \cdot \nabla f_i$, in Eq. (11) cannot be homogenous. Also, the second sound is the temporal and spatial fluctuation of temperature field which requires the description under unsteady condition. Therefore, we would need to solve the PBE as an original form containing both differential and integral terms. We briefly review the past approaches several decades ago to solve the PBE with several assumptions, and also introduce recent approaches with minimal assumptions from first principles.

One of the most challenging parts of solving the PBE is how to handle the integral scattering term. In the PBE, all phonon states are coupled to each other through the integral scattering term. Callaway suggested a simple form of scattering model from the fact that N-scattering and U-scattering tend to relax a phonon system to displaced and stationary Bose-Einstein distributions, respectively [43]. Although the Callaway's scattering model was from intuition without rigorous theoretical considerations, it was later shown that the model can be formally derived by ignoring the off diagonal terms of the N- and U-scattering matrices [44].

Early theoretical studies of phonon hydrodynamics derived macroscopic transport equations like the Navier-Stokes equation of fluid flow [4, 5, 7, 8, 32]. The work by Sussmann and Thellung [4] solved the PBE to the first order assuming no U-scattering and constructed momentum and energy balance equations. Some of Krumhansl group's work extended the transport equations to the case where U-scattering exists [5, 6]. The notable work by Guyer and Krumhansl [7, 8] solved the PBE in the eigenstate space of scattering operator which led to the concept of relaxon that will be discussed later. The derivation of these early studies were carefully examined and compared later by Hardy [32]. Although the details of derivation in the early studies are slightly different, they share the same basic idea. The idea is similar to how the Navier-Stokes equation is derived from the Boltzmann transport equation with the BGK scattering model which is analogous to the N-scattering term of the Callaway's scattering model. We briefly discuss the Sussmann and Thellung's derivation here.

The momentum and energy balance equations can be simply derived from the PBE by taking momentum ($\hbar \mathbf{q}$) and energy ($\hbar \omega$) as a moment of the PBE:

$$\frac{\partial E}{\partial t} + \nabla_\alpha Q_\alpha = 0 \tag{13}$$

$$\frac{\partial P_\alpha}{\partial t} + \nabla_\beta p_{\alpha\beta} = 0 \tag{14}$$

where

$$E = \frac{1}{NV} \sum_i \hbar \omega_i f_i \tag{15}$$

$$Q_\alpha = \frac{1}{NV} \sum_i \hbar \omega_i v_{\alpha,i} f_i \tag{16}$$

$$P_\alpha = \frac{1}{NV} \sum_i \hbar q_{\alpha,i} f_i \tag{17}$$

$$p_{\alpha\beta} = \frac{1}{NV} \sum_i \hbar q_{\alpha,i} v_{\beta,i} f_i \tag{18}$$

The $E$ and $Q_\alpha$ are the energy density and heat flux along the $\alpha$-direction. The $P_\alpha$ and $p_{\alpha\beta}$ are the $\alpha$-direction momentum density and the momentum flux along $\beta$-direction. Note that the right-hand side of Eq. (13) and Eq. (14) are zero because total momentum and energy are conserved upon N-scattering. If U-scattering is considered, the momentum destroying term by U-scattering would appear in the momentum balance equation. In order to complete those momentum and energy balance equations, the phonon distribution function is required. The phonon distribution can be found by solving the PBE with the N-scattering term of Callaway's scattering model:

$$\frac{\partial f}{\partial t} + \mathbf{v} \cdot \nabla f = -\frac{f - f^{disp}}{\tau_N} \tag{19}$$

Eq. (19) can be further simplified if we assume $\dot{f} \approx \dot{f}^{disp}$ and $\nabla f \approx \nabla f^{disp}$. This assumption is analogous to the Chapman-Enskog expansion to the first order and is valid when N-scattering is strong [45]. To be more specific, N-scattering is considered strong when the relaxation time and mean free path of N-scattering are much smaller than the characteristic time and size of system (e.g., time period of temperature fluctuation for second sound and the sample size for steady-state heat flow). With such assumptions, it is straightforward to solve Eq. (19). Based on the phonon distribution function from Eq. (19) being plugged into Eq. (13) and Eq. (14), the following macroscopic governing equations can be derived:

$$\dot{T}' - \frac{1}{3}v_g^2 \nabla^2 T' + \frac{1}{3}\nabla \cdot \mathbf{u} = 0 \qquad (20)$$

$$\dot{u}_\alpha + v_g^2 \nabla_\alpha T' - v_g^2 \tau_N \left(\frac{2}{5}\nabla_\alpha \nabla \cdot \mathbf{u} + \frac{1}{5}\nabla^2 u_\alpha\right) = 0 \qquad (21)$$

where $v_g$ is the group velocity. The $T'$ is the dimensionless deviational temperature defined as $(T - T_0)/T_0$ where $T_0$ is an equilibrium temperature.

Although early theoretical studies [4, 5, 7, 8, 32] are slightly different in the details of derivation, they are based on the same assumptions: i) N-scattering being much stronger than U-scattering such that $f$ is closer to $f^{disp}$ than $f^0$, and ii) N-scattering being strong enough that $\dot{f} \approx \dot{f}^{disp}$ and $\nabla f \approx \nabla f^{disp}$. Because of these assumptions, the hydrodynamic equations derived in the early studies have several limitations. The macroscopic hydrodynamic equations may not accurately describe the following cases: i) the characteristic size of system being comparable to the mean free path of N-scattering, namely phonon transport in somewhere between ballistic and hydrodynamic limits, and ii) N-scattering being not much stronger than U-scattering, namely phonon transport in somewhere between diffusive and hydrodynamic limits. In addition, the validity of Callaway's scattering model is questionable for quantitative purposes [46, 47].

As the full scattering matrix can now be calculated from first principles and the hydrodynamic phonon transport gained the renewed attention, there are two recently developed methods to solve the PBE with the full scattering matrix in both real and reciprocal spaces without the assumption of strong N-scattering. Both approaches provide a solution of the PBE without any significant assumptions and thus can be useful to study a complex transport phenomena where features of all three regimes exist to some extent [48].

The first approach is based on the eigenstates of the scattering matrix. The scattering matrix can be symmetrized by multiplying a factor, $2\sinh(X_i/2)$ where $X_i = \hbar\omega_i/k_B T$ to Eq. (11) such that the scattering matrix has an orthogonal set of eigenstates [44]:

$$\left(2\sinh\frac{1}{2}X_i\right)\mathbf{v}_i \cdot \nabla f_i = \sum_j G_{ij}^* f_j^{d*} \qquad (22)$$

where $f_j^{d*}$ is $\left(2\sinh\frac{1}{2}X_j\right)f_j^d$ and the scattering matrix, $\mathbf{G}^*$, is

$$G_{ij}^* = \left( \frac{2\sinh\frac{1}{2}X_i}{2\sinh\frac{1}{2}X_j} \right) G_{ij} \tag{23}$$

The orthogonal eigenstates of $\mathbf{G}^*$ are later called relaxons [49]. The solution of the PBE, $f^{d*}$, then can be expressed as a linear combination of relaxons and the equation for the coefficient (population) of each relaxon state can be derived from the PBE [49]. An advantage of the relaxon framework is that relaxon has now a well-defined relaxation length and thus the thermal transport can be described with a simple kinetic description of relaxon particles. Phonons, if experience strong N-scattering, do not have well-defined relaxation length due to the complex interplay between N- and U-scattering processes and also its collective nature of motions.

The second approach employs the Monte Carlo (MC) method to solve the PBE with the full scattering matrix [50, 51]. The MC method was previously developed to solve the PBE with the single mode relaxation time approximation (SMRT) for studying quasi-ballistic phonon transport [52-54]. The MC method with the SMRT stochastically determines the occurrence of scattering based on the probability of scattering. With the full scattering matrix, the MC method stochastically determines whether a certain scattering process occurs or not and the final state of phonon particles if the scattering is determined to occur. The energy-based PBE is chosen over the regular PBE due to its advantage of strict energy conservation:

$$\mathbf{v}_i \cdot \nabla (\omega_i f)_i = \sum_j B_{ij} (\omega_j f_j^d) \tag{24}$$

where $B_{ij}$ is the scattering matrix of the energy-based PBE, defined as $(\omega_i/\omega_j) G_{ij}$. The energy exchange upon scattering is described as

$$\omega_i f_i^d (t+\Delta t) = \sum_j Z_{ij}(\Delta t) \omega_j f_j^d (t) \tag{25}$$

where the energy propagator matrix $\mathbf{Z}$ can be found as

$$\mathbf{Z}(\Delta t) = e^{\mathbf{B}\Delta t} \tag{26}$$

If the off-diagonal terms of matrix $\mathbf{B}$ is ignored and only diagonal terms are considered, Eq. (25) is recovered to the exponential decay of energy which is equivalent to the SMRT:

$$\omega_i f_i^d (t+\Delta t) = \exp(B_{ii}\Delta t) \omega_i f_i^d (t) \tag{27}$$

Note $B_{ii}$ is the same as $-\tau_i^{-1}$ from Eq. (24). The scattering in Eq. (25) describes the transfer of energy from phonon state $j$ to $i$. In MC simulation, the destination state $i$ can be stochastically determined and its detailed MC algorithm can be found in literatures [50, 51].

**IV. Steady-State Phonon Hydrodynamics**

The N-scattering itself does not directly cause thermal resistance because of its momentum conserving nature. However, the N-scattering can affect thermal resistance when combined with momentum-destroying scattering (R-scattering or diffuse boundary scattering) or thermal reservoirs that emit phonons of which distribution deviates from the displaced Bose-Einstein distribution. These situations are common in practical systems. We discuss the role of N-scattering for thermal resistance in three different cases: i) an infinitely large sample, ii) a sample with an infinite length but a finite width where diffuse boundary scattering destroys the phonon momentum along the flow direction, and iii) a sample with an infinite width but a finite length contacting hot and cold reservoirs that emit phonons with the stationary Bose-Einstein distribution.

IV.1. Infinitely Large Sample

It is well known that the thermal conductivity is infinitely large when the N-scattering is the only scattering mechanism and the sample is infinitely large. Assuming that a local temperature gradient is applied and phonon flow is initiated, the phonons subsequently establish the displaced Bose-Einstein distribution through many N-scattering events. Then, the N-scattering does not further alter the displaced Bose-Einstein distribution and the phonons can continue to flow even without any temperature gradient, resulting in the infinite thermal conductivity. This leads to the simple statement that N-scattering itself does not cause thermal resistance. This simple statement, however, is true only when the distribution function is homogenous in space as in the infinitely large sample. If there is a significant spatial variation of the distribution function, the N-scattering can cause thermal resistance. This will be discussed in IV.3.

Even when the distribution function is homogenous in space, N-scattering contributes to thermal resistance if U-scattering also exists. In general, phonon states with a small wavevector have very weak U-scattering. However, the small wavevector phonons can be scattered into larger wavevector states through N-scattering and then can be seen by U-scattering. A recent study on the thermal transport in SWCNTs [55] quantitatively shows the effect of N-scattering on thermal conductivity. The thermal conductivity of (10,10) SWCNT is 10,000 W/m-K when only U-scattering is considered, but it is significantly reduced to 2,000 W/m-K when both N- and U-scattering processes are included.

IV.2. Sample with an Infinite Length and a Finite Width

We consider a sample with an infinite length and a finite width to discuss the thermal resistance when N-scattering is combined with diffuse boundary scattering. As shown in Fig. 6, we consider a constant temperature gradient along the length direction, which drives the phonon flow.

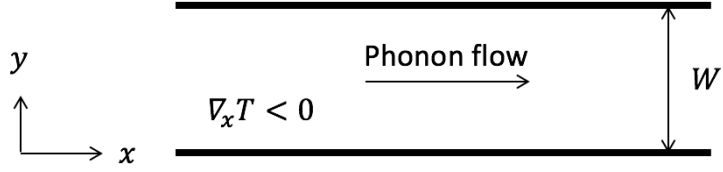

Fig. 6. Schematic of phonon flow in an infinitely long sample with a finite width

The major mechanisms of thermal resistance in diffusive and ballistic regimes are U-scattering and diffuse boundary scattering, respectively. In the hydrodynamic regime, we have a different mechanism for thermal resistance: viscous damping effect which is a result of combined N- and diffuse boundary scattering. The drift velocity near boundaries is smaller than that in the middle of a sample due to diffuse boundary scattering. Thus, the drift velocity exhibits a gradient along the transverse direction ($y$-direction in Fig. 6). Due to the drift velocity gradient, phonon momentum is transferred from the middle of the sample to the boundaries through N-scattering processes and then finally is destroyed by the diffuse boundary scattering. The viscous damping term can be seen in the second-order derivative term in Eq. (21).

Based on the momentum balance equation from the PBE with its first-order solution discussed in section III, an expression for the phonon hydrodynamic viscosity ($\mu_{ph}$) can be derived [51]:

$$\mu_{ph} = \frac{\sum_i q_x^2 v_{y,i}^2 f_i^0 \left(f_i^0 + 1\right) \tau_{N,i}}{\sum_i q_x v_{x,i} f_i^0 \left(f_i^0 + 1\right) \omega_i} \tag{28}$$

A notable difference between ballistic and hydrodynamic regimes is that the momentum transfer to the boundary in hydrodynamic regime is impeded by N-scattering. As N-scattering rate is increased, the rate of momentum transfer to the boundary, which determines the extent of viscous damping, is decreased. This can be seen in the phonon hydrodynamic viscosity as a function of temperature in Fig. 7. As temperature increases, the N-scattering rate is increased, resulting in the lower hydrodynamic phonon viscosity. The extent of viscous damping also depends on the width of sample as indicated in the second-order derivative term in Eq. (21). The rate of momentum transfer rate in hydrodynamic regime is proportional to $1/W^2$ where $W$ is the width of a sample, while the rate in ballistic regime is proportional to $1/W$.

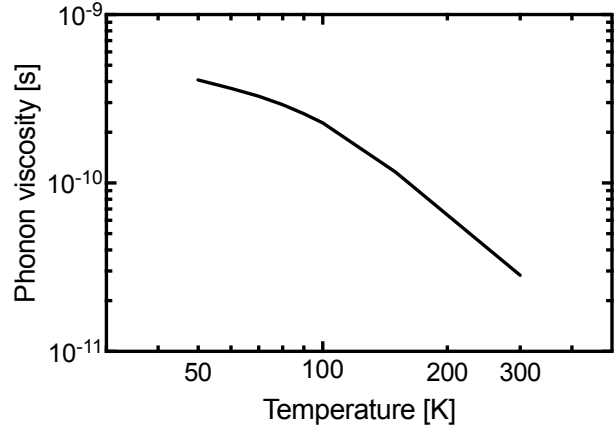

Fig. 7. Temperature dependence of phonon hydrodynamic viscosity of suspended graphene calculated with phonon dispersion and scattering rates from first-principles calculation [51].

The viscous damping effect of hydrodynamic regime causes peculiar dependences of thermal conductivity on temperature and sample width, which are distinguished from the ballistic and diffusive cases. In the ballistic regime, the thermal conductivity is linearly proportional to the sample width. The thermal conductivity of diffusive regime is constant regardless of a sample width. However, the thermal conductivity of hydrodynamic regime superlinearly increases with the sample width due to the viscous damping that decreases as $W^2$. In addition, the thermal conductivity of hydrodynamic regime increases with temperature much faster than that of ballistic regime as the viscous damping is weakened as temperature increases. The peculiar dependence of thermal conductivity on temperature was observed in solid He at low temperature, verifying the existence of phonon Poiseuille flow [11]. Recently these dependences have been predicted at much higher temperature in graphene [3, 51, 56] and graphite [20], and experimentally observed in $SrTiO_3$ [57].

The peculiar dependences of thermal conductivity on temperature and sample width can be observed only when the actual transport phenomena are close to those in the ideal hydrodynamic regime without U-scattering. The thermal transport in graphitic materials at intermediate temperature above 100 K can exhibit all three different mechanisms of thermal resistance: U-scattering, direct diffuse boundary scattering, combined diffuse boundary and N-scattering. The significance of each mechanism can be evaluated using the momentum balance. The temperature gradient in Fig. 6 drives phonon flow and generates excess phonon momentum ($\Phi_{\nabla T}$). This momentum is balanced by momentum destructions by three different mechanisms: diffuse boundary scattering without internal phonon scattering (i.e, ballistic effect, $\Phi_B$), diffuse boundary scattering combined with N-scattering (i.e., viscous damping or hydrodynamic effect, $\Phi_H$), and direct momentum destruction by U-scattering (i.e., diffusive effect, $\Phi_D$). The momentum balance can be expressed as

$$\Phi_{\nabla T} = \Phi_B + \Phi_H + \Phi_D \tag{29}$$

Fig. 8(a) shows that thermal conductivity of an infinitely long graphene has a temperature dependence of $T^{2.03}$ when temperature is below 90 K, much larger than that of the ballistic case $T^{1.68}$. This temperature range agrees with the momentum balance analysis of the same sample, shown in Fig. 8(b). It is clear that below 90 K, $\Phi_H$ is the major mechanism of the momentum destruction, indicating that viscous damping is significant at this condition [48].

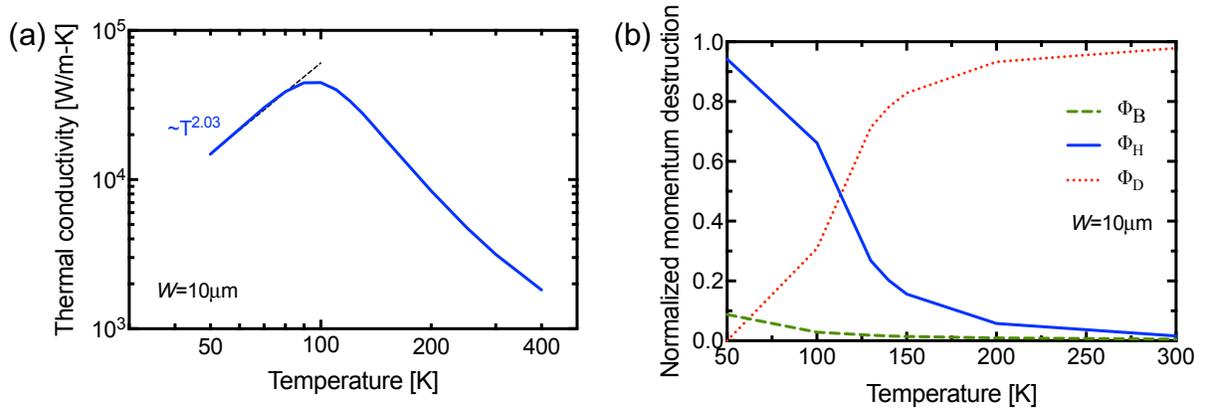

Fig. 8 Temperature dependence of (a) thermal conductivity, and (b) the momentum balance in an infinitely long graphene sample with the width of 10 μm from the MC solution of the PBE with *ab initio* full three-phonon scattering matrix.

IV.3. Sample with an Infinite Width and a Finite Length Contacting Hot and Cold Reservoirs

When an infinitely wide sample contacts hot and cold reservoirs as in Fig. 9, the phonons emitted from the reservoirs do not follow the displaced Bose-Einstein distribution. They follow a Bose-Einstein distribution distorted by a spectral transmission function at the interface between the sample and the reservoir. The N-scattering processes change this non-displaced Bose-Einstein distribution (i.e., non-collective) to the displaced Bose-Einstein distribution (i.e., collective). As entropy is always generated when the distribution function is changed by scattering processes as shown in Eq. (6), N-scattering causes thermal resistance near the interface between the sample and the reservoir where the emitted phonon flow becomes collective. The region where the thermal resistance occurs is within the order of the mean free path of N-scattering from the boundary. Fig. 10 shows the formation of collective phonon flow at the cost of temperature drop near the boundaries, resulting in the thermal resistance by N-scattering. Assuming N-scattering is the only scattering mechanism, the N-scattering far from the boundaries does not cause any temperature drop as the distribution function is already the displaced Bose-Einstein distribution.

The thermal resistance due to the transition between non-collective and collective phonon flows depends on materials. Fig. 11 compares three-dimensional Debye phonon dispersion and graphite in terms of the reduction of phonon heat flux by N-scattering from the purely ballistic case. For the 3D Debye case, the reduction of heat flux is relatively small; the heat flux reduction by N-scattering is only around 5% for all three temperatures, 100, 200, and 300 K. However, for graphite, the reduction of heat flux is substantial; the heat flux is reduced by 20, 30, and 40% at 100, 200, and 300 K, respectively. Other graphitic materials such as SWCNTs and graphene show similar reduction of heat flux.

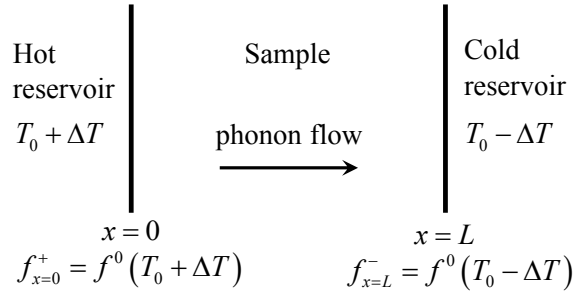

Fig. 9 Schematic of sample geometry contacting hot and cold reservoirs

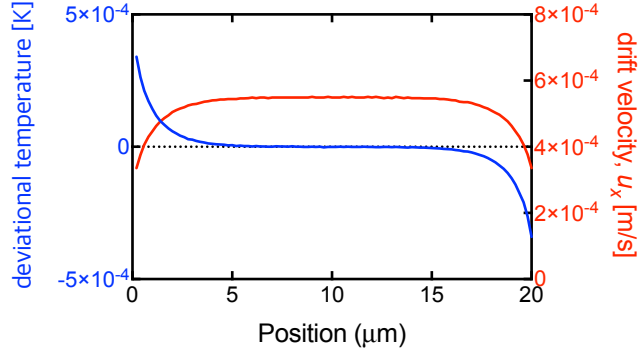

Fig. 10 The profile of deviational temperature, defined as the difference between local temperature and global equilibrium temperature, and the drift velocity. The sample is (20,20) SWCNT contacting hot and cold reservoirs that have the deviational temperature of 0.001 and -0.001 K, respectively. The profile is calculated by Monte Carlo method of the PBE assuming the Callaway's scattering model. The rate of N-scattering is assumed $10^{10}$ s$^{-1}$ and U-scattering is ignored [58].

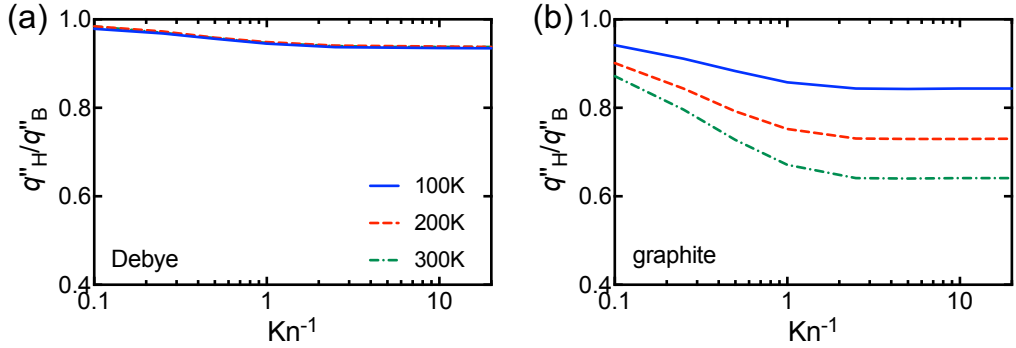

Fig. 11 The ratio between heat flux with N-scattering ($q''_H$) and without any internal scattering ($q''_B$) as a function of inverse Knudsen number in (a) three-dimensional Debye model and (b) graphite. The heat flux is calculated with the Monte Carlo solution of the PBE with the Callaway's scattering model. The rate of N-scattering is assumed $10^{10}$ s$^{-1}$ and U-scattering is ignored [58].

The large thermal resistance by N-scattering for graphitic materials can be explained with their non-linear phonon dispersion with many phonon branches. The rate of entropy generation due to scattering, Eq. (6) can be written as follows assuming the Callaway's scattering model and the stationary Bose-Einstein distribution for phonons emitted from the reservoirs.

$$\dot{S}_{scatt} = \left(\frac{\Delta T}{T}\right)^2 \frac{\hbar^2}{\tau_N k_B T^2 NV} \sum_i f_i^0 \left(f_i^0 + 1\right) \omega_i \left|q_{x,i}\right| \left(v^*_{x,i} - u'_x\right) \qquad (30)$$

where $v^*_{x,i}$ is $\omega_i / \left|q_{x,i}\right|$ and $u'_x$ is the drift velocity per temperature difference and can be found from the momentum conservation:

$$u'_x = \frac{\sum_i |q_{x,i}| \omega_i f_i^0 (f_i^0 + 1)}{\sum_i q_{x,i}^2 f_i^0 (f_i^0 + 1)} \quad (31)$$

The Eq. (30) shows that the temperature difference of two reservoirs drives the phonon flow with the displacement of $v^*_{x,i}$ which may vary depending on phonon modes while the drift velocity $u'_x$ is the same for all phonon modes. If $v^*_{x,i}$ is a constant for all phonon modes (e.g., one-dimensional Debye phonon dispersion), $v^*_{x,i}$ is the same as $u'_x$ by Eq. (31) and the entropy generation would be zero. However, if $v^*_{x,i}$ significantly varies with phonon states, the entropy generation is expected to be large. For the three-dimensional Debye model, the $v^*_{x,i}$ varies with the direction of phonon wavevector and thus causes small thermal resistance as can be seen in Fig. 11(a). The ratio $q''_H / q''_B$ in this case does not change with temperature as the variance of $v^*_{x,i}$ is associated with the direction of phonon wavevector only. For graphic materials, however, the variance of $v^*_{x,i}$ is significant compared to the Debye model as a result of non-linear dispersion with many branches, resulting in the large thermal resistance that depends on temperature in Fig. 11(b). This indicates the resistance due to the transition between non-collective and collective phonon flows is determined by the shape of phonon dispersion.

## V. Unsteady Phonon Hydrodynamics (Second Sound)

The fundamental difference between N- and U-scattering in terms of momentum conservation leads to the different response upon temporal perturbation to a phonon system. One simple form of the perturbation is a heat pulse being applied to one end of a sample as shown in Fig. 12. The heat pulse causes the increased local phonon density and the response of phonon system is largely different depending on the transport regime. For the diffusive regime, the energy balance equation with the Fourier's law indicates that the peak position of heat pulse cannot move forward and remains at its original location. Then, the thermal energy of the heat pulse diffuse into the sample and finally the sample reaches an equilibrium with a slightly elevated temperature for the entire region. For the ballistic regime, the heat pulse can propagate through the sample as there is no phonon scattering. However, the shape of heat pulse can spread out in space unless all phonon modes have the same group velocity along the heat pulse propagation direction. This is expected to be particularly significant in graphitic materials where flexural phonon modes with a quadratic dispersion are important for thermal energy transport. For the hydrodynamic regime, the heat pulse leads to the local fluctuation of temperature field which can propagate as a wave through the sample. An analogous phenomenon in fluid system is the propagation of pressure pulse in space, which is acoustic sound. From the similarity of the two phenomena, the temperature pulse propagation in the form of wave in hydrodynamic regime is called second sound.

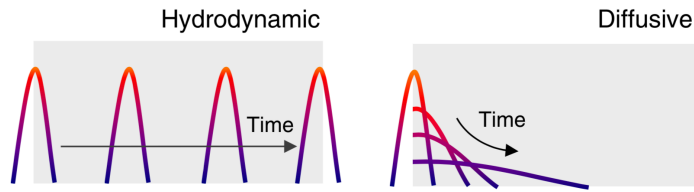

Fig. 12. Propagation of a heat pulse in diffusive and hydrodynamic regimes [3]

The second sound was first studied with superfluid He II in which phonon is an elemental excitation of the system. The speed of second sound in liquid He was predicted by Landau using the two fluid theory of rotons and phonons [59] and later confirmed by an experiment [60]. The predicted temperature wave was named as second sound by Landau in order to distinguish it from the first sound which is the ordinary acoustic sound (i.e., pressure wave propagation). Later it was shown that the same speed of second sound can be directly derived by using phonon gas model without roton [61, 62], which motivated the studies of second sound in crystalline solids.

The second sound in solids can be observed with two different methods: heat-pulse experiment [12-15, 63] and light scattering method [10, 64-69]. In the heat pulse experiment, a heat pulse was applied to one end of a few mm long sample and the temperature to the opposite end was recorded as a function of time. At sufficiently low temperature such that internal phonon-phonon scattering is negligibly weak, two peaks of temperature pulse were observed, each of which represents the ballistic transport of transverse and longitudinal phonons. No significant dispersion of the temperature peak was observed because three dimensional bulk materials where long wavelength phonons have a linear dispersion relation were used. With slightly increased temperature (around 15 K for NaF [14]), another peak in addition to those two peaks were observed. The delay time of the new peak agree well with the predicted speed of second sound and the third peak was considered second sound. As temperature is further increased, the third peak disappears, indicating that U-scattering becomes significant. The light scattering method measures the inelastic light scattering by a local change of dielectric constants due to second sound wave. A challenge lies in very weak coupling between light and thermal fluctuation at low temperatures. To solve this problem, a relatively strong thermal fluctuation field was induced by an optical grating method and the second sound in NaF was successfully measured [67]. The measured speed of second sound agrees well with that from the previous heat pulse experiments. Later, the light scattering measurements were carried out without inducing thermal fluctuation field for $SrTiO_3$. The $SrTiO_3$ has soft transverse optical phonons with small wavevector that are strongly anharmonic and thus cause strong N-scattering [68, 69]. The measured spectrum at around 30 K exhibits a doublet with a frequency shift (~ 20 GHz) that is comparable to the expected frequency of second sound in this temperature range.

As the conditions for the clear observation of second sound is narrow in the variable space, the second sound measurements critically require *a priori* knowledge on the wavelength and frequency of second sound as well as the speed of second sound. The wavelength and frequency of second sound are determined by the mean free path and scattering rate of N- and U-scattering processes. If the pulse duration is much longer than the rate of U-scattering, the pulse can be destroyed by the U-scattering and thus cannot propagate as a second sound. If the pulse duration is much shorter than the rate of N-scattering, phonons will travel with their own group velocity and do not have a chance to establish the collective motion because of the lack of N-scattering. In this case, the pulse also cannot maintain its original shape and the thermal energy smears out.

The speed, frequency, and wavelength of second sound were theoretically studied by calculating the dispersion relation of second sound. The speed of second sound was derived for the simplest case where Debye phonon dispersion is assumed and there is no U-scattering, giving the well-known relation for the speed of second sound, $v_{II} = v_g/\sqrt{3}$, where $v_{II}$ is the speed of second sound [61, 62]. The speed of second sound was also derived for more realistic phonon dispersion consisting of one longitudinal and two degenerate transverse acoustic branches, all having the Debye-type dispersion [4]. Later theoretical studies considered the possible mechanisms for attenuation of second sound and predicted the possible second sound frequency ranges [5-7]. In the literature, two different types of second sound called drifting and driftless second sounds were discussed [22, 70]. The driftless second sound differs from the second sound we discuss and does not require strong N-scattering; it occurs when all eigenstates of scattering operator have the similar relaxation time such that collective-looking thermal transport can occur. To our best knowledge, there was no experimental observation of the driftless second sound.

The dispersion relation of second sound can be derived from the momentum and energy balance equations in Eq. (13) and Eq. (14). If U-scattering is considered, the momentum destruction by U-scattering needs to be added to the right hand side of Eq. (14). An example dispersion relation of second sound in (20,20) SWCNT is shown in Fig. 13(a). The real and imaginary frequencies represent the propagation and attenuation of a pulse, respectively. The imaginary frequency in the limit of small wavevector (i.e., long wavelength) is mostly determined by the rate of U-scattering. As the wavevector is increased (i.e., wavelength becomes shorter), the viscous damping effect by N-scattering becomes strong, causing the significant attenuation of second sound. Fig. 13(b) shows the required length of a sample. For the second sound to propagate, the sample length should be larger than the wavelength of second sound but smaller than its relaxation length defined as $v_{II} \text{Im}(\Omega)$ where $v_{II}$ and $\text{Im}(\Omega)$ are the speed of second sound and the imaginary part of second sound frequency, respectively.

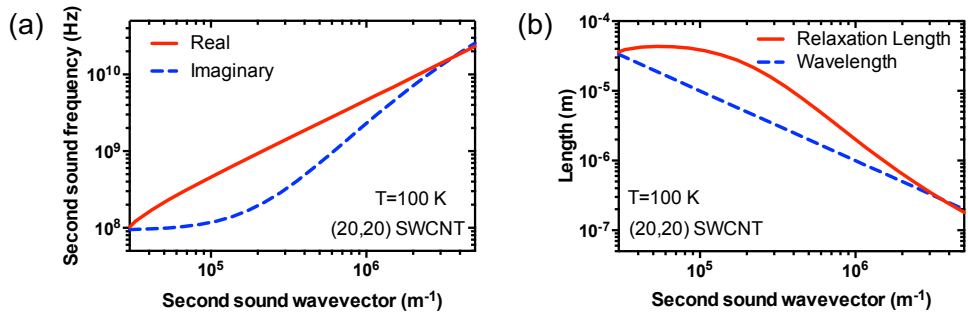

Fig. 13. The propagation and attenuation of second sound in (20,20) SWCNT. (a) the dispersion relation of second sound showing propagation (real) and attenuation (imaginary) of second sound. (b) the comparison between relaxation length and wavelength of second sound. [18]

## VI. Summary and Future Perspectives

In this chapter, we briefly reviewed the past and recent studies on hydrodynamic phonon transport. We first discussed the displaced Bose-Einstein distribution representing the collective motion of phonon particles as an equilibrium state under N-scattering. Then, we introduced several approaches to solve the Peierls-Boltzmann transport equation for the case where N-scattering is significant. Based on the solution of the Peierls-Boltzmann transport equation, we then showed how N-scattering affects thermal phonon transport in both steady state and transient cases. For the steady state cases, we discuss three-scenarios; when N-scattering is combined with i) U-scattering, ii) diffuse boundary scattering, and iii) thermal reservoirs that emit phonons following non-displaced Bose-Einstein distribution functions. In all cases, N-scattering affects thermal transport indirectly. For the first case where N-scattering is combined with U-scattering, it transfers energy from small wavevector states where U-scattering is relatively weak to large wavevector states where U-scattering is strong, thereby contributing to thermal resistance. For the second case, the N-scattering impedes the momentum transfer to the boundaries which acts as a momentum sink by diffuse boundary scattering. We discussed that stronger N-scattering leads to less viscous damping effect and larger thermal conductivity. For the last case, the N-scattering itself causes thermal resistance when the distribution function is not homogenous in space due to thermal reservoirs emitting phonons with non-displaced distribution. The thermal resistance occurs while those non-collective phonon flows become collective through N-scattering processes. The thermal resistance by the transition between collective and non-collective phonon flows depends on the shape of phonon dispersion; while the thermal resistance due to this effect is small for Debye phonon dispersion, it can be significant in graphitic materials because of their highly non-linear phonon dispersion with many branches. The second sound was discussed as a representative phenomenon of phonon hydrodynamics in transient case. The N-scattering causes the damping of second sound even without U-scattering. If the fluctuation of temperature field is fast in time and space domains, for example the frequency and wavelength of second sound are shorter than the rate and mean free path of N-scattering respectively, the fluctuation can be largely damped. Thus, the N-scattering imposes the limit of frequency and wavelength of second sound for its propagation.

Although the recently developed *ab initio* framework for phonon transport has been proved for its high accuracy and predictive power [42], the significant hydrodynamic phonon transport in graphitic materials need to be experimentally confirmed. The significant contributions from the flexural phonon modes to thermal transport were experimentally shown [71], but its strong N-scattering due to extremely large anharmonicity for small wavevector states has not been verified. The explicit observation of hydrodynamic phonon transport has several challenges. First, the measurements need to be done in much smaller length and time scale compared to the previous studies performed several decades ago. The characteristic length and time scale of hydrodynamic phonon transport scales with the mean free paths and rate of internal phonon scattering. As those previous studies measured the hydrodynamic phonon transport at extremely low temperature below 15 K, the internal phonon scattering was weak; therefore the Poiseuille flow was measured with several mm size sample [11] and the second sound propagation was measured with the time scale of μs [14]. However, as the hydrodynamic phonon transport in graphitic materials is expected to occur at much higher temperature, the internal phonon scattering is accordingly strong. Thus, the experiments need to be done with sub-mm size samples and ns temporal resolution. Recent advancements on the microscale platform for the measurement of thermal conductivity [72, 73] as well as the ultrafast spectroscopy technique [74-76] are perhaps well suited for the measurement of hydrodynamic phonon transport in graphitic materials. Second, we would need a large sample with minimal defects. The *ab initio* simulation shows that the sample size should be at least 10 μm for measuring phonon Poiseuille flow and second sound at 100 K [51], but typical graphitic material samples with this sample size contain many defects. Interestingly, the observation of second sound was reported very recently using a highly oriented pyrolytic graphite (HOPG) sample [77]. This study used the transient grating method to generate the standing wave of second sound and could measure the fluctuation of temperature which is associated with the second sound. As the second sound is in standing-wave form in this study, it does not need to propagate throughout the entire sample and could be measured with less significant damping. The observation of phonon Poiseuille flow is expected to be more challenging compared to the second sound case. The theoretical prediction of phonon Poiseuille flow assumed infinitely long samples for the condition of fully developed phonon flow [51]. If a sample has a finite length, there would be so called entrance effect which is due to the transition from spatially uniform phonon flow to parabolic phonon flow near the entrance. This would require a sample with length being much larger than width. In the previous study, the phonon Poiseuille flow was predicted with the width of 10 μm, thus length should be much longer than this value.

    The recent prediction of significant hydrodynamic phonon transport indicates that the hydrodynamic regime is practically important for high thermal conductivity materials where N-scattering is often strong and cannot be ignored. Although the clear observation of hydrodynamic phonon transport is expected at sub-room temperatures, the hydrodynamic phonon transport is still important for understanding the thermal transport. As shown in Fig. 3, the mean free path of N-scattering and U-scattering have a large gap in the length ranges from sub-μm to μm for 300 K. If the sample size lies in this gap which is common in the practical applications of high thermal conductivity materials for thermal management, the diffusive-ballistic phonon transport may not correctly describe the thermal transport phenomena; it ignores the thermal resistance due to the momentum transfer and formation of collective phonon flow by N-scattering. Therefore, the hydrodynamic regime needs to be considered another limit of thermal transport in addition to ballistic and diffusive limits which were extensively studied in the past [78, 79]. The detailed mechanisms of how N-scattering contributes to thermal resistance when combined with other scattering processes have not been rigorously discussed in the past. We think this is partially because of the lack of available numerical tools; it has been very challenging to solve the Peierls-Boltzmann transport equation in both real and reciprocal spaces with minimal assumptions. With the recently developed *ab initio* frameworks for solving the Peierls-Boltzmann transport equation in both real and reciprocal spaces [49, 51], it is now possible to quantitatively study the influence of N-scattering on the overall thermal transport process when it is combined with other scattering

processes. This would complete the understanding of phonon transport in high thermal conductivity materials and lead to the better design of thermal devices using those high thermal conductivity materials.

**Acknowledgement**

We acknowledge support from National Science Foundation (Award No. 1705756 and 1709307).